\documentstyle[aps,epsf,rotate,subfigure,epsfig]{revtex}

\newcommand{\eeq}{\end{equation}}
\newcommand{\beq}{\begin{equation}}

\newcommand{\ba}{\begin{array}}
\newcommand{\ea}{\end{array}}
\newcommand{\veps}{\varepsilon}

\begin{document}
\draft
\twocolumn[\hsize\textwidth\columnwidth\hsize\csname @twocolumnfalse\endcsname

\title {Multifractal Properties of Return Time Statistics}
\author{Nicolai Hadyn} \address{Dept. of Mathematics,
University of Southern California, Los Angeles},
\author {Jos\'{e} Luevano} \address{ Centre de Physique Theorique,
Univ. de Aix Marseille II, and Universidad Autonoma Metropolitana,
Azcapotzalco, Mexico},
\author {Giorgio Mantica}
\address{International Center for the Study of Dynamical Systems,
Universit\`a dell'Insubria, Via Valleggio 11, Como, Italy, and
I.N.F.M., unit\`a di Milano, I.N.F.N. sez. Milano,
giorgio.mantica@uninsubria.it}
\author
{Sandro Vaienti} \address{Centre de Physique Th\'eorique, Luminy,
Marseille and PHYMAT, Universit\'e de Toulon et du Var, France,
and F\'ed\'eration de Recherche des Unit\'es de Math\'ematiques
de Marseille}
\date{\today}
\maketitle
\begin{abstract}
Fluctuations in the return time statistics of a dynamical system
can be described by a new spectrum of dimensions. Comparison with
the usual multifractal analysis of measures is presented, and
difference between the two corresponding sets of dimensions is
established. Theoretical analysis and numerical examples of
dynamical systems in the class of Iterated Functions are
presented.
\end{abstract}
 ]

Suppose that $T$ is a transformation of the space $X$ into itself,
which preserves the probability measure $\mu$. Let $A$ be a subset
of $X$, and $x$ a point in $A$. We let $\tau_A(x)$ be the
(integer) time of the first return of $x$ in $A$:
\begin{equation}
 \tau_A(x) = \inf \{ n>0 \; \mbox{s.t.} \;  T^n(x)\in A \} .
\label{rito1}
\end{equation}

Poincar\'e theorem guarantees that
the return time is almost certainly finite with respect to any
invariant measure. It can be rather long,
and particularly short as well: if $x$ is a fixed point of $T$,
then obviously $\tau_A(x) = 1$ for any set $A$ containing $x$. In
this paper, we shall prove that the distribution of return times
is characterized by multi-fractal properties which can be properly
described  by the tools of the so-called thermodynamical formalism
\cite{ruelle}.

Much recent research has been focused on the {\em local}
statistics of these returns, in the following sense: Take for $A$
a ball of variable radius $\veps$ centered at 
the point $x$: $A= B_\veps(x)$.  Next,
consider the cumulative distribution of the first
return times, in the ball, of all points of this latter:
Let $\lambda(\veps) = \mu (B_\veps(x))$, and define
 $  m(x,\veps,s) := {\mu \{ y \in B_\veps(x) \;
\mbox{s.t.} \;
    \lambda(\veps)  \tau_{B_\veps(x)}(y) > s \}} / \lambda(\veps) $.
In many instances, these statistics have an exponential character:
 $ m(x,\veps,s) \rightarrow e^{-  s}$, as $\veps$
tends to zero.  Rigorous proofs
of this fact have been produced under an ever lessening set of
hypotheses \cite{vostriamici}.

More fundamental for our analysis than the previous, is Kac
theorem \cite{Kac}, a classical result of the local analysis: it
predicts that, whenever the measure $\mu$ is ergodic with respect
to $T$, for any measurable set $A$ (and not just balls), the
expectation of $\tau_A$ over  the set $A$ is just the inverse of
the measure of $A$:
\begin{equation}
  \int_A \tau_A(y) \frac{d\mu(y)}{\mu(A)} = \frac{1}{\mu(A)}.
 \label{kac2}
 \end{equation}

Furthermore, one must quote the theorem of Ornstein and Weiss
\cite{Ornstein-Weiss}: Let $\cal A$ be  a generating partition of
$X$, and let us refine it around all the points $x \in X$ by
defining the cylinder of order $n$, $A_n(x)$, as the intersection
of all the elements of $\cal A$, $T^{-1}\cal A$,...,$T^{-n+1}\cal
A$ containing $x$. The Ornstein-Weiss theorem states that,
whenever the measure $\mu$ is ergodic, the following limit exists
$\mu$-almost everywhere, and is equal to $h(\mu)$, the metric
entropy of $\mu$:
\begin{equation}
\lim_{n\rightarrow\infty}\frac{\log \tau_{A_n(x)}(x)}{n} = h(\mu).
\label{owe}
\end{equation}

This remarkable result is historically  the first link between a
thermodynamic quantity, and the return times. Parallel to the
Ornstein-Weiss theorem, the Shannon-Mc Millan-Breiman theorem
gives the metric entropy by means of the exponential decay of the
measure of cylinders around almost all points. It is common in the
physical usage to replace cylinders with balls: this allowed in
force of the Brin-Katok theorem \cite{Brin-Katok}. The advantage
of balls over cylinders, when considering practical applications,
is apparent.

Let us therefore replace $A_n(x)$ and  $n$ in eq. (\ref{owe}) by
$B_\veps(x)$ and $- \log \veps$, respectively.  In the case of
Gibbs measures of Axiom-A diffeomorphisms \cite{Barreira-Saussol},
 and of a wide class of maps of the interval
\cite{Saussol-Troub-Vaienti} one has been able to prove rigorously
that the modified limit (\ref{owe}) exists $\mu$-almost
everywhere, and is equal to the {\em information dimension}
$D_\mu(1)$,  of $\mu$:
 \begin{equation}
     \lim_{\veps \rightarrow 0} - \frac{\log(\tau_{B_\veps(x)}(x))}{
     \log \veps }= D_\mu(1).
 \label{sand1}
 \end{equation}
The definition of $D_\mu(1)$, and an informal proof of eq.
(\ref{sand1}), will be given momentarily.

So far, the analysis has been local. Quite different--and more
complex--is the case of the {\em global} statistics that we
consider in this paper. We let balls of fixed radius $\veps$ cover
$X$, and we compute  the integrals
 \begin{equation}
 \Gamma_\tau(\veps,q) := \int_X \tau_{B_\veps(x)}^{1-q}(x)
 d\mu(x),
 \label{nte1}
 \end{equation}
where $q$ is a real quantity. These are the statistical moments of
the time required to come back to a neighborhood of the starting
point: it is clear that local analysis alone has little to say
about the scaling, in the radius $\veps$, of these quantities.

Indeed, $\Gamma_\tau(\veps,q)$ are a sort of {\em partition
function}, quite akin to those employed in the thermodynamical
formalism. Their scaling for small $\veps$ defines a set of
dimensions, $D_\tau(q)$:
 \begin{equation}
 \Gamma_\tau(\veps,q) \sim \veps^{D_\tau(q)(q-1)},
 \label{nte3}
 \end{equation}
that we call {\em return time dimensions}. The study of these
dimensions is the object of this Letter. In level of importance,
the first question is about their relations with the usual
quantities of the thermodynamical formalism.

 Let us first observe
that the return time $\tau_{B_\veps(x)}(x)$ can be interpreted as
a ``single-point'' sample of the integral in Kac theorem, eq.
(\ref{kac2}). Therefore, one could estimate that
 \begin{equation}
\tau_{B_\veps(x)}(x) \sim \mu(B_\veps(x))^{-1}.
  \label{esti}
 \end{equation}
It is worth recalling now that the {\em local dimension}
$\alpha(x)$ of the measure $\mu$ at the point $x$ is defined by
the scaling $\mu(B_\veps(x)) \sim \veps^{\alpha(x)}$, in the limit
of small $\veps$. Moreover, the relation $\alpha(x) =
D_\mu(1)$ holds almost everywhere in $X$, as discussed above. This
fact, and eq. (\ref{esti}), then imply the theorem expressed by
eq. (\ref{sand1}).

The approximate equality (\ref{esti}) has been already adopted in
\cite{grass,chica} to evaluate  {\em via} eq. (\ref{nte1}), the
{\em exact} thermodynamical partition function
$\Gamma_\mu(\veps,q)$,
 \begin{equation}
 \Gamma_\mu(\veps,q) := \int_X \mu({B_\veps(x)})^{q-1}
 d\mu(x) \sim \veps^{D_\mu(q)(q-1)},
 \label{nte2}
 \end{equation}
whose scaling for small $\veps$ gives the well-known spectrum of
generalized dimensions $D_\mu(q)$ \cite{pesin book}.

The substitution of eqs. (\ref{nte2}) by eqs.
(\ref{nte1},\ref{nte3}) is particularly significant,  for it
allows one to treat dynamical systems endowed with physical
measures of the Sinai-Bowen-Ruelle type: in this case the integral
can be replaced by a Birkhoff sum over the trajectory $x_l :=
T^l(x_0)$, $l=0,\ldots$, of a generic point $x_0$:
 \begin{equation}
 \Gamma_\tau(\veps,q) = \lim_{n \rightarrow \infty}
  \frac{1}{n} \sum_{j=0}^{n-1}
 \tau_{B_\veps(x_j)}^{1-q}(x_j).
 \label{nte4}
 \end{equation}
Indeed, this Birkhoff procedure, eq. (\ref{nte4}), is the one
originally employed in \cite{grass,chica}, where it is claimed to
produce the spectrum of measure dimensions $D_\mu(q)$.

It must be remarked that eq. (\ref{nte4}) considers a single sum
along the orbit, while in the usual Grassberger-Procaccia
technique for the $q$-correlation integrals, one must work with a
double summation, where all the couples of points along the same
(or different) orbit are compared \cite{Grassberger-Procaccia}.

We are now ready to introduce the main result of this paper:
contrary to the usage of \cite{grass,chica}, the estimate
(\ref{esti}) does hold only for the first moment, as in Kac
theorem, but not in the stronger sense required in eqs.
(\ref{nte1}), and (\ref{nte2}): the variables $\tau_{B_\veps(x)}$
and $\mu(B_\veps(x))^{-1}$ have different large deviation
properties. Therefore, $D_\mu(q)$ {\em and} $D_\tau(q)$ {\em are
not the same function}, and {\em the latter defines a new bona
fide spectrum of dimensions}.

The abstract proof of this statement would be little informative,
if not paralleled by a specific example. It is therefore
convenient  to introduce a family of dynamical systems whose
invariant measures are completely known, in the sense that the
spectrum of dimensions $D_\mu(q)$ can be easily and precisely
computed: these are the so-called systems of iterated functions,
or I.F.S., see \cite{barn} for details. In the simplest,
one-dimensional case, an I.F.S. consists of a collection of
contracting maps of the line, $\phi_i : {\bf R} \rightarrow {\bf
R}$, $i=1,\ldots,M$, where $M$ is the number of maps. These latter
can be thought of as the inverse branches of the dynamics: $(T
\circ \phi_i) (x) = x$ for any $i$. In so doing, $T$ is
characterized by a mixing repeller, that is also the attractor of
the collective action (in the sense that will be made soon clear)
of the maps in the I.F.S..

A family of invariant measures for $T$ can be constructed
assigning a {\em probability}, $\pi_i$, to each map $\phi_i$,
% $i =1,\ldots,M$,
$\sum_{i=1}^M \pi_i = 1$. The dynamics of I.F.S. can be
constructed by sequentially applying maps $\phi_\sigma$, where
$\sigma$ is chosen in a random fashion with probability
$\pi_\sigma$. Any orbit of the probabilistic I.F.S., when
time-reversed, becomes an orbit of the deterministic dynamics $T$.
Notice that return times are invariant under time reversal.

The reader will find useful to consider the usual ternary Cantor
set measure as the invariant measure of the following 2-maps
I.F.S.: $\phi_1(x) = x/3$, $\phi_2(x) = (x+2)/3$,
$\pi_1=\pi_2=1/2$. The related invariant measure is an example of
{\em mono-fractal}: all generalized dimensions are equal to the
constant $\delta = \frac{\log 2}{\log 3}$. Let us perform the
return time analysis of this dynamical system following the
Birkhoff sums approach, eq. (\ref{nte4}). Other techniques of
estimating the integral (\ref{nte1}) will be presented elsewhere.

\begin{figure}[htbp]
\centering\epsfig{file=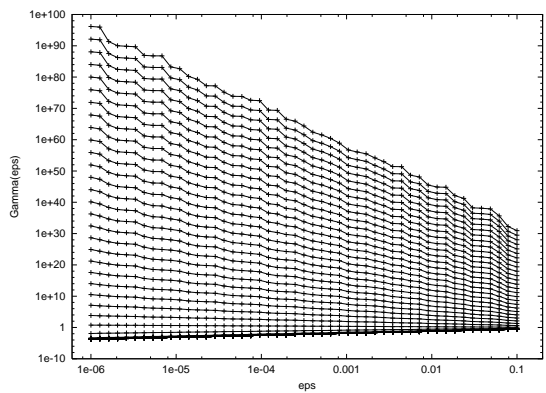,width=0.6\linewidth,angle=270}
\caption{ Return time partition sums $\Gamma_\tau(\veps,q)$ versus
$\veps$ for a set of equally spaced values of $q$ ranging from
$q=-19$ (bottom curve) to $q=19$ (top). Lines in between points
are solely intended to connect data with the same value of $q$. }
\protect\label{gamma}
\end{figure}

We can first verify that the scaling relation (\ref{nte3}) holds:
fig. \ref{gamma} assures us that this is indeed the case.

\begin{figure}[htbp]
\centering\epsfig{file=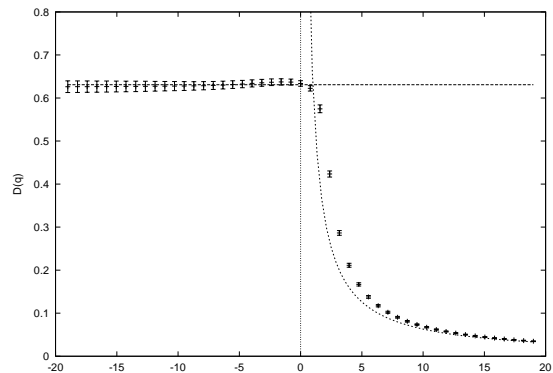,width=0.6\linewidth,angle=270}
\caption{Generalized dimensions $D_\tau(q)$ versus $q$ for the
ternary Cantor set dynamics. Error bars are defined via the
uncertainty in the least square fits of Fig. (\ref{gamma}), under
the usual statistical assumptions.  The horizontal line is the
(flat) measure thermodynamics : $D_\mu(q) = \log 2 / \log 3$. The
dashed curve for $q>0$ is eq. (\ref{nte10}) with $\alpha = \log 2
/ \log 3$. } \protect\label{termo}
\end{figure}

We can then extract the {\em return} dimensions $D_\tau(q)$ versus
$q$: the numerical results are plotted in fig. \ref{termo}. They
reveal a non-trivial range of dimensions: as far as return times
are concerned, the dynamics is multifractal. Moreover, the return
time dimensions are consistent with the constant value $\delta$
(within the statistical error bars) for {\em negative }
dimensions, but are significantly different from this latter for
{\em positive} values of $q$, the more, the larger the value of
$q$.

Resorting now to exact analysis, we can confirm these results:
precisely, we are able to prove that {\em i)} $D_\mu(0) =
D_\tau(0)$, and {\em ii)} $D_\tau(q) \rightarrow 0$ for large $q$.
The second fact proves that the return dimensions $D_\tau(q)$ are
significantly different from measure dimensions, while the first
hints at relations that, at least in certain dynamical systems,
may exist between them.

To see that {\em i)} is true, observe that the repeller of $T$ can
be hierarchically covered by intervals $A^k_l$, of length $1/3^k$,
where $k=0,\ldots$ is the index of the generation, and
$l=1,\ldots,2^k$ is the label of the single interval within its
generation. This is simply the usual construction of the ternary
Cantor set, in which one carves a {\em hole} of length $1/3^k$ in
the middle of each interval generation $k-1$.

For a fixed value of $k$, let us choose $\veps = \veps_k = 1/3^k$.
Then, simple geometric considerations assure us that whenever $x$
belongs to $A^k_l$, with $k$ and $l$ fixed, it is also contained
in the ball of radius $\veps_k$ centered around any other point in
$A^k_l$. In addition, these balls intersect no  interval of
generation $k$ other than $A^k_l$. This permits us to employ Kac
theorem, eq. (\ref{kac2}), to compute {\em exactly} the sum
$\Gamma_\tau(\veps_k,0)$:
 \begin{equation}
 \Gamma_\tau(\veps_k,0)
 % = \int_X \tau_{B_\veps_k(x)}^{1-q}(x) d\mu(x)
  = \sum_{l=0}^{2^k} \int_{A^k_l} \tau_{A^k_l}(x) d\mu(x) =
  2^k.
 \label{nte8}
 \end{equation}
It is then clear that {\em i)} holds.

Remark that the intervals $A^k_l$, are also the cylinders of order
$k$ of the dynamics.  Therefore, the result we have just proven
also applies to the return time dimensions, when defined via
cylinders, and predicts that the dimension of order zero is equal
to the topological entropy. Following the same idea, one also
expects that the cylinder return time dimensions will be related
to the Renyi entropies of the measure  \cite{nota}.

Observe that the above results are by no means restricted to the
one-dimensional case. At the same time it is evident that use of
Kac  theorem is permitted only for $q=0$, for otherwise different
fluctuation properties of the return time statistics set in. This
leads us directly to the statement {\em ii)}.

Remark that the return times define partitions of $X$ by the sets
$R_n(\veps) := \{ x \in X \; \mbox{s.t.} \; \tau_{B_\veps(x)}(x) =
n \}$. This allows us to re-write the moments
$\Gamma_\tau(\veps,q)$ as a new summation:
 \begin{equation}
 \Gamma_\tau(\veps,q) = \sum_{n=1}^{\infty} n^{1-q}
 \mu(R_n(\veps)).
 \label{nte7}
 \end{equation}
When $q$ is large, terms with small $n$ lead the sum (\ref{nte7}).
Let us retain just the first of these: it does not vanish by
increasing $q$. This term is the measure of $R_1(\veps)$, the set
of points that move less than $\veps$ in a single iteration.
Clearly, all fixed points of $T$ belong to this set. Let $\bar{x}$
be any one of these, and let us assume that $\veps$ is small
enough, and that the transformation $T$ is smooth. Then,
$R_1(\veps)$ contains the ball of radius $ \veps /\Lambda  $
centered at $\bar{x}$, where $\Lambda$ is the largest singular
value of the matrix ${\bf 1} - T'$,  ${\bf 1}$ is the identity,
and $T'$ is the Jacobian matrix of $T$ at $\bar{x}$. Moreover, the
ball of radius $ \veps / \sigma$, where $\sigma$ is now the
smallest singular value, contains the connected part of
$R_1(\veps)$ around $\bar{x}$. Then, we can estimate that, for
large $q$,
\begin{equation}
 \Gamma_\tau(\veps,q) \simeq \sum_{\bar{x} \mbox{ s.t. }
 T(\bar{x}) = \bar{x}}  \mu(B_{\rho \veps} (\bar{x})),
 \label{nte9}
 \end{equation}
where the sum is extended to all fixed points of $T$,
and where $\rho$ is a multiplicative factor, which depends on
$\bar{x}$:
$\sigma < \rho < \Lambda$. When the number of terms in the
sum is finite, the leading contribution to (\ref{nte9}) comes
from the fixed point $\bar{x}$ with the {\em smallest} local
dimension, $\alpha_m$: since $\mu(B_\veps (\bar{x})) \sim
\veps^{\alpha_m}$, it follows at once that
\begin{equation}
 D_\tau(q) \simeq \frac{\alpha_m}{q-1}, \; \mbox{when} \; q
 \rightarrow \infty.
 \label{nte10}
 \end{equation}

We can readily verify that eq. (\ref{nte10}) is satisfied by the
Cantor set dynamics introduced above. The fixed points of $T$ are
here zero, and one, and they are both characterized by the same
value of the local dimension, $\alpha_m = \frac{\log 2}{\log 3}$.
In fig. \ref{termo} the relation (\ref{nte10}) is validated by the
numerical results of $D_\tau(q)$ for large $q$.

\begin{figure}[htbp]
\centering\epsfig{file=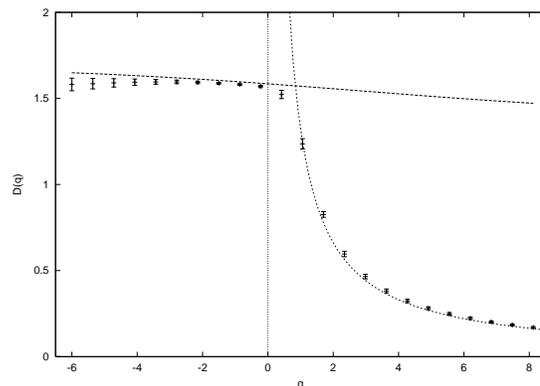,width=0.6\linewidth,angle=270}
\caption{Generalized dimensions $D_\mu(q)$ (dotted line) and
$D_\tau(q)$ (error bars) versus $q$ for the Sierpinsky dynamics
described in the text. Data have been computed by least square
fits of Birkhoff sums with $n=800000$. Also reported for positive
$q$ is the asymptotic formula (\ref{nte10}). }
\protect\label{sierp}
\end{figure}

This investigation is not limited to a single dimension. We now
add a two-dimensional example with a 
non-trivial dimension function $D_\mu(q)$: the motion on a Sierpinsky gasket,
corresponding to the I.F.S. maps $\phi_1(x,y) = (x/2,y/2)$,
$\phi_2(x,y) = ((x+1)/2,(y+1)/2)$, and $\phi_3(x,y) =
(x/2,(y+1)/2)$, with non-uniform probabilities $\pi_1= 4/10$,
$\pi_2=\pi_3= 3/10$. In fig. \ref{sierp} we have compared the
exact thermodynamical function $D_\mu(q)$ and the numerically
evaluated $D_\tau(q)$. For negative values of $q$ the two
dimensions are very close, and discrepancies may be due to the
finite length of the Birkhoff sum, and, more importantly, of the
fitting interval in $\veps$, which cannot be accounted for by the
statistical definition of the error bars. For positive values of
$q$, to the contrary, the two dimensions are rather different, and
the asymptotic formula (\ref{nte10}) soon becomes a very close
approximation of $D_\tau(q)$.

In conclusion, we have shown that the statistics of return times
are characterized by multi-fractal properties, well defined by a
{\em new} spectrum of dimensions $D_\tau(q)$. We have also found
that, contrary to implicit previous statements in the literature,
this spectrum does {\em not} coincide with the usual multi-fractal
spectrum $D_\mu(q)$. It is now interesting to understand the
meaning of the Legendre transforms associated to these new
dimensions. We have proven that $D_\tau(0)=D_\mu(0)$ under certain
hypotheses on the dynamical system. Approximate equality seems to
hold for negative $q$'s as well \cite{nota2}. This might provide a
means for computing negative measure dimensions, which is known to
be a challenging numerical task. Finally, we have shown that in a
large class of systems the relation $D_\tau(q) \simeq
\frac{\alpha_m}{q-1}$ holds for large $q$, where $\alpha_m$ is the
smallest local dimension at the fixed points of $T$.

\begin{thebibliography}{99}
\bibitem{nota} If one considers cylinders in place of balls,
the two generalized dimensions can be identified with the
deviation functions of the
variables $\tau_{A_n(x)}(x)$ and $\mu({A_n(x)}(x))$ \cite{colgas}.
Then, one can
rigorously prove that the two dimensions coincide in a
small neighbourhood to the right of $q=1$, but they must differ for large $q$
\cite{hv}.
\end{thebibliography}
\end{document}